\documentclass[11pt]{article}
\usepackage{template}
\usepackage{tikz}
\usetikzlibrary{patterns}
\usepackage{authblk}
\begin{document}
\title{\rule{\textwidth}{3pt}\\\vspace{-0.7em}\rule{\textwidth}{2pt} Online variable-weight scheduling with preempting on jobs with linear and exponential penalties\\\rule{\textwidth}{2pt}\vspace{-0.7em}\\\rule{\textwidth}{3pt}}
\author{Frederick Tang \footnote{ftang@mit.edu} \hspace{2em} Fareed Sheriff\footnote{fareeds@mit.edu} \hspace{2em} Andrew Wang\footnote{andrewva@mit.edu}\\\hspace{0.5em}MIT\hspace{7.5em} MIT\hspace{6.5em} MIT}
\maketitle
\begin{abstract}
\par We analyze the problem of job scheduling with preempting on weighted jobs that can have either linear or exponential penalties. We review relevant literature on the problem and create and describe a few online algorithms that perform competitively with the optimal scheduler. We first describe a na{\" i}ve algorithm, which yields a high competitive ratio ($\Omega(\frac{M}{s_{\min}})$) with the optimal, then provide an algorithm that yields a lower competitive ratio ($4\sqrt{\frac{M}{s_{\min}}} + n\log{\frac{Mn}{s_{\min}}}$). Finally, we make a minor modification to our algorithm to yield an algorithm that has an even better competitive ratio ($n\log{\frac{Mn}{s_{\min}}}$).
\end{abstract}
\newpage
\section{Introduction}
Job scheduling has been widely studied and in many flavors, and it involves scheduling a set of $n$ jobs, each of which take a certain time to process, to be processed on 1 or more machines where each machine can only process a single job at a time. The goal is to minimize the total penalty incurred, which usually takes the form $\sum_{i=1}^n C_iw_i$ where $C_i$ is the time job $i$ finished processing and $w_i$ is the penalty weight associated with job $i$. Job scheduling has many similarities to operating system kernels. An OS kernel on one processor/machine is given processes/tasks to complete and must complete these jobs sequentially while prioritizing some tasks over others. Some processes, like a mouse click or keyboard press, need to be completed quickly, so we assign them high penalty weights, and the processor can optimize scheduling processes by minimizing the total penalty. Kernels do not know the future, so they have to allocate jobs to the processor online.
\par Many job scheduling variants fit this model, but they are oversimplified; processes like mouse clicks do not have to be done immediately, and computers should be allowed to wait some amount of cycles (perhaps a few million). However, once we reach a certain time threshold, it becomes very bad for the mouse click to not be processed yet: taking 0.2 seconds to process is not the end of the world, but after waiting 1 second, taking another 0.2 seconds will make users very frustrated. Thus, linear penalties like end time multiplied by $w_i$ cannot capture this. We introduce the problem of exponential penalties $x^{C_i}$ for base $x$; After a few time units, the penalty is not that large, but each time unit the process remains unscheduled exponentially increases its penalty, encouraging kernels to process it soon. We also allow kernels to pause a process and switch to more urgent processes; this is known as preempting. We propose an algorithm for this problem: scheduling jobs with either linear or exponential penalties on a single machine with preempting.
\section{Related Works}
\subsection{Online Job Scheduling on a Single Machine with General Cost Functions}
\par This paper \cite{9682957} describes an algorithm for online job scheduling with preempting using arbitrary penalty functions. Every job is associated with a nondecreasing cost function, completion time, and processing length. The authors analyze their algorithm's competitiveness and speed using speed augmentation, which allows comparison of the online algorithm's performance on a machine that runs at unit speed to an optimal scheduling on a machine that runs at speed $\frac{1}{1+\eps},\eps>0$. Rather than expressing job costs as the product of a weight $w_j$ and completion time $C_j$, this weight and other characteristics of the job are collected into a nondecreasing cost function $g_j(C_j)$.
\par $v_j$ is defined to be processing length of job $j$ with $v_j(t)$ denoting the fraction of the job uncompleted at time $t$, $r_j$ is the release time of job $j$, and $g_j(\cdot)$ is the cost function of $j$, with $C_j$ being the earliest time at which $j$ has been processed as normal. The problem to be solved is to minimize $\sum_j{g_j(C_j)}$ --- we want to minimize the sum of job costs under their respective cost functions.
The objective represents that we want to minimize the total cost of all jobs over the time they're processed. The first constraint ensures every job is completely processed while the second enforces a unit-time processing speed. The algorithm in \cite{9682957} aims to have an objective cost competitive with the LP objective above.
\par The authors find such an algorithm by first analyzing the HGFC problem with identical release times, abbreviated HGFC($r_n$), describing how to solve it, further analyzing the problem through the lens of LPs, then describing an online algorithm for HGFC.
\par We can describe the time $r_n$ when job $n$ is released using $\cal{A}(r_n)$, which is the set of alive jobs including $n$ at time $r_n$, and note that an optimal scheduling for $\cal{A}(r_n)$ assuming no more jobs will be released is an optimal scheduling for the offline problem. The state of all jobs can be defined using a vector of residual lengths $\mathbf{v}(t) = \langle v_1(t),\cdots, v_k(t)\rangle$ where each $v_i(t)$ represents a job in $\cal{A}(t)$ and $v_i(t)$ is the residual length of the job at $t$.
\par The authors then use the minimum principle to solve HGFC($r_n$). The minimum principle is a principle in optimal control theory \cite{minprinciple} that says any optimal control plus the optimal state trajectory must satisfy the Hamiltonian. Basically, given a system of ordinary differential equations with goal to find an optimal control $\mathbf{u}^*$ and path to this control $\mathbf{x}^*$ minimizing some objective, the minimum principle says in order for $\mathbf{u}^*(t),\mathbf{x}^*(t)$ to be optimal, $\mathbf{u}^*$ must necessarily maximize the system Hamiltonian $H(\mathbf{u})$. The authors express the previous equation as an optimal control theory problem.
With the proper boundary conditions, this yields the following method for obtaining the optimal offline schedule of HGFC($r_n$): we can plot the job curves $p_k(t_1)-g_k(t)\forall k\in\cal{A}(r_n)$ and at time $t$, we schedule the job with the maximum value of $p_k(t_1)-g_k(t)$. To determine the actual values of $p_k(t_1)$, we find the dual of our LP and solve. We define $\beta_t = p_j(t_1)-g_j(t)$ if $j$ is scheduled at time $t$ and $\frac{\alpha_j}{v_j(r_n)} = p_j(t_1)$. Then, we see that $\beta_t + g_j(t) = p_j(t_1)-g_j(t)+g_j(t) = \frac{\alpha_j}{v_j(r_n)}\leftrightarrow \frac{\alpha_k}{v_k(r_n)} = \beta_t+g_k(t)$.
\par Finally, the authors use the Hamilton-Jacobi-Bellman equation \cite{hjb} (an equation in optimal control theory whose solution is an optimal value function that allows us to find values for the costate constants \cite{costate} $p_i$) to describe a competitive algorithm for the online HGFC problem.
\par In the online problem, the cost functions of job $j$ is released at $r_j$ and we want a scheduling that that processes jobs competitively with an optimal offline schedule. This online algorithm is to greedily schedule jobs based on the optimal offline policy of all current jobs --- at any moment, the optimal policy is to go with the schedule on the currently active jobs that minimizes the sum of costs over all jobs assuming no further jobs will be added to the set of currently active jobs (thus treating the current schedule as offline). The time complexity of this algorithm is $\O(nT_s)$ where $T_s$ is the complexity of solving HGFC($r_n$); this can be approximated in polytime with a $(1+\eps$ quasipolynomial algorithm or $(4+\eps)$ strongly-polynomial algorithm.
\par This greedy algorithm is analyzed in the paper using the dual of the offline LP, the concept of a monotone substitute cost function (a cost function that is nondecreasing differentiable and does not change previous optimal scheduling intervals except by shifting them after being added to the set of active jobs), and speed augmentation to show that the algorithm is $(1+\eps)$-speed 2-competitive for HGFC.
\subsection{List Scheduling in Order of $\alpha$-Points on a Single Machine}
\par Skutella\cite{Skutella2006-bk} surveys the landscape of list scheduling algorithms in a variety of different formulations: he considers scheduling using preemption, precedence, and release dates along with different combinations of all three. The author shows multiple methods and algorithms including using Smith's rule, the use of $\alpha$ points, and an LP relaxation. The paper first analyzes $1\|\sum_{j}{w_jC_j}$ and lays out Smith's rule or a job sorting using a heuristic, $t_j/w_j$ where $t_j$ and $w_j$ are the total processing time and penalty of job j, that is then proven to result in the optimal ordering. The paper then analyzes the problem, $1\|r_j\|\sum_{j}{C_j}$, where $r_j$ are the release dates of jobs or the earliest a job j can be scheduled and $C_j$ is the completion time.  The author shows an approximation scheme using alpha points. The author then shows a fractional preemptive schedule using a linear program that can serve as an upper bound for a preemptive schedule. The following section will explore these three methods in more depth.

The author first explores Smith’s rule: let job $i$ have processing time $p_j$ and linear penalty $p_j$; scheduling in order of increasing $t_j/w_j $ gives an optimal solution for $1||\sum w_jC_j$. This was proved via an exchange argument. Consider the following example. Say we have two jobs $i_1$ and $i_2$, and in our current schedule $i_2$ immediately follows $i_1$ but $\frac{t_{i_1}}{w_{i_1}} > \frac{t_{i_2}}{w_{i_2}}$. Rearranging, $t_{i_1}w_{i_2}> t_{i_2}w_{i_1}$. Say we swapped the two job positions. The only difference in total penalty from this swap is local or occurs in the changes of the penalties of jobs $i_1$ and $i_2$. So the difference in penalty after swapping is:
\begin{align*}
    w_{i_2}t_{i_2}+ w_{i_1}(t_{i_2}+t_{i_1}) -w_{i_1}t_{i_1}-w_{i_2}(t_{i_1}+t_{i_2}) &=w_{i_1}t_{i_2}-w_{i_2}t_{i_1}\\
    &<0 \ \ \ \ \ \ \ \ \ \ \ \ \ \text{by above}
\end{align*}
Thus, every time we swap two out of order jobs, we reduce the penalty. So the optimal penalty is when all jobs are ordered by increasing $\frac{t_i}{w_i}$. Note this is a generalization of SPT (shortest processing time) as for schedules with no weights, we schedule the job with the lowest $t_i$ first.\\

To find the online preemptive schedule with weights, we can maintain a priority queue sorted by the heuristic used in Smith's rule, $p_j/w_j$. We can pop jobs once they're finished and process the next job up at every time step. 

Smith concludes this heuristic is a 2-approximation for $1| r_j , pmtn | \sum w_jC_j$. This is shown by lower bounding the completion time by the sum of the release date added to the processing times of all the jobs that were initiated beforehand. If we let $C_j$ denote the completion time of job j in the preemptive list schedule, by definition, we have $C_j \leq r_j + \sum_{k \leq j} p_k$ where $p_k$ are the processing times of jobs that are before and including job j. In other words, the completion time of any job j is lower-bounded by both the release date and the processing times of all the jobs before and including job j. If we multiply both sides by $w_j$: 

\begin{equation*}
    \sum_{j}C_jw_j \leq \sum_{j}r_jw_j + \sum_{j}w_j\sum_{k \leq j}p_k
\end{equation*}

By definition, the optimal completion time is always after its release date so the optimal completion time for any job $j$ is lower bounded by $\sum_j w_jr_j$. The optimal completion time is also lower bounded by $\sum_j w_j * \sum_{k \leq j} p_k$ by Smiths' rule, if we imagine this as equivalent to having 0 release dates. This leads to these two lower bounds being both less than the optimal scheduling cost and thus, we have a 2-approximation. 

Skutella also considers another approach, using Alpha points. Alpha points are the point when $\alpha$-fraction amount of a job is completed where $\alpha \in (0, 1]$. As the author shows: for a fixed job j and $0 \leq \alpha \leq 1$ the $\alpha$-point $C^P_j(\alpha)$ of job j can be written as:

\begin{equation*}
    C^P_j(\alpha) = t_{idle}(\alpha) + \sum_{k \in J} \eta_k(\alpha)p_k \geq \sum_{k \in J} \eta_k(\alpha)p_k.
\end{equation*}

where $C^P_j(\alpha)$ is the completion time of the alpha point of job j and $\eta_k(\alpha)$ is the fraction of jobs k completed before time $C^P_j(\alpha)$.

Goemans et al.\cite{10.5555/314161.314394} furthers this idea by choosing $\alpha$ randomly and being job-dependent. The author shows this leads to an approximation for $1|r_j|\sum C_j$:


\begin{equation*}
    \E[\sum_j w_jC^{\alpha}_j] = \sum_j w_j\E[C^{\alpha}_j] \leq \dfrac{e}{e-1}\sum_j w_jC^{P}_j.
\end{equation*}

We  know how to solve $1|r_j, pmtn|\sum C_j$ optimally by processing in order of remaining processing time. This concludes an $\dfrac{e}{e-1}$ approximation for $1|r_j|\sum C_j$. The algorithm for finding $1|r_j|\sum C_j$ is as follows:
\begin{enumerate}
    \item create a preemptive schedule by processing in order of remaining processing time
    \item randomize a value for $\alpha$ from [0, 1] with density function $f(x) = \dfrac{e^x}{e-1}$
    \item process in order of completion time of $\alpha$ according to the preemption schedule.
\end{enumerate}
Dyer and Wolsey\cite{Dyer1990-fe} also lay out an LP relaxation to the non-preemptive problem $1|r_j, pmtn|\sum w_jC_j$ that can be extended to the precedence case. They introduce an indicator variable $x_{j, t}$ which is set to 1 if job j starts processing at time t and 0 otherwise. The given LP is:
\begin{align*}
    \min{\sum_{j \in J} {w_jC_j^{LP}}}, \quad&\\\st \sum_{t = r_j} ^{T}{x_{jt}} = 1 \quad&\forall j,\\ \sum_{j \in J} \sum_{l = max(0, t-p_j+1)}^{t}{y_{jl}} \leq 1 \quad&\forall t = 0, ..., T,\\ \sum_{l = r_j}^{t}{x_{jl}} \geq \sum_{l = r_k}^{t+p_j}{x_{kl}} \quad&\forall j \prec k, t = r_j, ..., T-p_j,\\ C_j^{LP} = p_j + \sum_{t = r_j} ^{T}{x_{jt}} \quad&\forall j \in J,\\ x_{jt} = 0 \quad&\forall j \in J, t = 0, ..., r_j-1,\\ x_{jt} \geq 0 \quad&\forall j \in J, t = r_j, ..., T
\end{align*}


Goeman et al.\cite{10.5555/314161.314394} continued by showing a randomized 2-approximation for $1|r_j, pmtn|\sum w_jC_j$. The key idea is to randomize $\alpha$ for each job. The algorithm is as follows:
\begin{enumerate}
    \item Construct the preemptive list schedule P in order of non-decreasing ratios $p_j/w_j$.
    \item For each job $j \in J$, randomly assign $a_j$ from [0, 1].
    \item Construct the resulting $\alpha$ schedule.
\end{enumerate}
This can be shown in analysis to be a randomized 2-approximation using the expected value of $\E[\sum_j w_jC^{\alpha}_j]$. In fact, it can be shown to be a 1.6853 randomized approximation using a different analysis.
\section{Description \& Analysis}
We present an algorithm for minimum weighted flow time, which means the time for completion of job $i$, $C_i$, is measured as the difference between its actual completion time and its release time, $r_i$. Most research done to solve weighted flow time (as opposed to actual completion time) use the technique of speed augmentation to provide a competitive algorithm against the optimal running on an algorithm with low speed; section 2.1's paper is an example. We instead present a competitive 1-speed algorithm. We also slightly modify the problem. All exponential penalties have the same base $x\geq 2$. However, we will allow different starting penalties $s_i$ for each job with an exponential penalty. For time $t$, we let $s'_i=s_ix_i^{t-r_i}$ be the current penalty for job $i$ where $r_i$ is the release time of $i$. We define $t_i$ to be the processing time of job $i$; if $i$ has been processed for some amount of time already, $t_i$ reflects this and equals the amount of processing time left for job $i$.
\subsection{Na{\" i}ve Algorithm}
\par We first propose a na{\" i}ve algorithm for this problem, and show that it is at least $\frac{M}{s_{\min}}$-competitive with the optimal where $M$ is the maximum linear penalty and $s_{\min}$ is the minimal starting penalty for any exponential job. Then, we introduce a novel algorithm and show it is at most $\left(\sqrt{\frac{M}{s_{\min}}}+n\log({\frac{Mn}{s_{\min}}}))\right)$-competitive. We then make a final modification to achieve a $n\log({\frac{Mn}{s_{\min}}})$ algorithm. We let linear penalty job $j$ have linear penalty $w_i$.
\par The na{\" i}ve approach is to always process the greatest penalty active job --- for every timestep $t$, we calculate for all active linear jobs the greatest $w_i$ and for the exponential jobs the greatest $s_i'$. We schedule the job corresponding to the larger of the two. We argue this algorithm is $\Omega(\frac{M}{s_{\min}})$-competitive. We provide an example of a case for which this algorithm is $\frac{M}{s_{\min}}$-competitive.
\begin{proof}
    \par Let there be a linear job $i$ with processing time $\log{\frac{M}{s_{\min}}}$ and penalty $M$. Consider an exponential job $j$ with $s_j=s_{\min}$ but large processing time $t\gg \log{\frac{M}{s_{\min}}}$. The na{\" i}ve algorithm will process $i$ until the penalty $s_j'$ becomes greater than $M$, which takes $\log{\frac{M}{s_{\min}}}$ time. The algorithm will then process the exponential job until completion. This results in the na{\" i}ve algorithm processing the linear job then the exponential job, and yields a penalty of $M\log{\frac{M}{s_{\min}}} + s_{\min}x^{t+\log{\frac{M}{s_{\min}}}} = M\log{\frac{M}{s_{\min}}} + Mx^t$. If we had instead completed job $j$ then job $i$, we would incur penalty $s_{\min}x^t + M(t + \log{\frac{M}{s_{\min}}}) = M\log{\frac{M}{s_{\min}}} + s_{\min}x^t + Mt$. This yields a competitive ratio of
    \begin{equation*}
        \frac{M\log{\frac{M}{s_{\min}}} + Mx^t}{M\log{\frac{M}{s_{\min}}} + s_{\min}x^t + Mt} = \frac{\log{\frac{M}{s_{\min}}} + x^t}{\log{\frac{M}{s_{\min}}} + t + \sfrac{s_{\min}x^t}{M}}\overset{t\rightarrow\infty}{\longrightarrow} \frac{M}{s_{\min}}\in\Omega{(\frac{M}{s_{\min}})}
    \end{equation*}
    so asymptotically, the na{\" i}ve algorithm is at least $\frac{M}{s_{\min}}$-competitive, making it $\Omega(\frac{M}{s_{\min}})$-competitive --- that is, it is at least $\frac{M}{s_{\min}}$-competitive but could do worse for some other example.
\end{proof}

\par We now refer to Skutella \cite{Skutella2006-bk}. As mentioned, Smith's rule says we sort all penalties by increasing density $\sfrac{t_i}{w_i}$. In the offline case where all jobs are released at the beginning, this is optimal.
\par We can use Smith's rule for exponential penalty jobs. Consider two exponential jobs $i, j$ where $i$ is scheduled before $j$. We want to find when it is optimal to swap the order of the two jobs to reduce the total penalty. This can be written
\begin{align*}
    s_ix^{t_i} + s_jx^{t_i+t_j} &> s_ix^{t_i + t_j}+s_jx^{t_j} \\\Longrightarrow s_ix^{t_i}(s_jx^{t_j}-1)&<s_jx^{t_j}(s_jx^{t_i}-1) \\\Longrightarrow \frac{s_ix^{t_i}}{s_jx^{t_i}-1} &< \frac{s_jx^{t_j}}{s_ix^{t_j}-1}
\end{align*}
This means we can sort by decreasing $\frac{s_ix^{t_i}}{x^{t_i}-1}$ to get the optimal offline scheduling. When processing online, we maintain active jobs in a priority queue sorted by $\sfrac{t_i}{w_i}$ where we process the top of the queue, preempting and switching to the top job as necessary.\\
However, we note that the analysis of Smith's rule in Skutella was for weighted completion time. Our goal is weighted flow time, since we only want to consider the penalty starting from when the job was released. So we cannot use the same analysis:\\

If we try to apply a similar analysis from Smith's rule's approximation of linear job scheduling with now instead cost measured from the release date:

\begin{equation*}
    \sum_{j}C_jw_j -  \sum_{j}r_jw_j \leq \sum_{j}r_jw_j + \sum_{j}w_j\sum_{k \leq j}p_k - \sum_{j}r_jw_j  
\end{equation*}

\begin{equation*}
    \sum_{j}w_j(C_j-r_j) \leq \sum_{j}w_j\sum_{k \leq j}p_k
\end{equation*}

\begin{equation*}
    \sum_{j}w_j(C_j-r_j) \leq \sum_{j}w_j\sum_{k \leq j}p_k
\end{equation*}

But if we bound the competitiveness, we achieve a $\sum_j C_j/(C_j-r_j)$ competitiveness. Thus, Smith's rule is not applicable to our problem.
\subsection{A Better Competitive Algorithm}
We will attempt to do well using a different sorting strategy. First, we need to change our priority queue ordering of the linear and exponential jobs. At each point in time, for every exponential job $i$, calculate the ending penalty of the job if we were to schedule that job to completion at that moment. This can be calculated by $s'_ix^{t_i}$, where $s'_i$ is the current penalty of the job and $t_i$ is the amount of processing time still remaining; we call $s'_ix^{t_i}$ the "potential" of job $i$. Let $i'$ be the job with the greatest potential. Store all linear jobs by decreasing weight. Then, when comparing exponential jobs with linear jobs, schedule $i$ before the top linear job $j$ in the queue if $s'_{i'}x^{t_{i'}}>\sqrt{\frac{M}{s_{\min}}}$, otherwise schedule the linear jobs. Note this modifies our problem to be partial-online, since we are need to be given beforehand $M$ and $s_{min}$.
\par At each timestep, if the exponential queue is nonempty, we process the top of the exponential queue; otherwise, we process the top of the linear queue. We now analyze the competitiveness of this algorithm against the penalty $C_{OPT}$ of a scheduling that minimizes the penalty sum where our algorithm's penalty for job $i$ is $P^i$ and $M=\max_{i\in L}{w_i}$ the maximum linear penalty weight.
\par Let $E_{OPT}$ be the optimal penalty if we only consider exponential penalties and ignore linear penalties; with individual optimal penalty $E^i_{OPT}$. Let also $L_{OPT}$ be the optimal penalty if we only consider linear penalties and ignore exponential penalties with individual optimal penalty $L^i_{OPT}$. It must be true that $C_{OPT}\geq E_{OPT}+L_{OPT}$ since the optimal of all jobs cannot be less than the sum of the optimal considering only exponential jobs and the optimal considering only linear jobs (we can meet equality with $C_{OPT}$ by running the assignment from $C_{OPT}$ on either exponential or linear jobs). We find a competitive bound by comparing our algorithm's exponential scheduling to the optimal exponential scheduling and likewise with linear jobs. For any job $i$, our algorithm could schedule $i$ suboptimally by either (1) scheduling a set of exponential jobs before $i$ where the optimal schedules the set after $i$ or (2) scheduling a set of linear jobs before $i$ but the optimal schedules the set after $i$. We can extend this to four cases:
\begin{enumerate}[(1)]
    \item Case (1) above where $i$ is exponential (exponential-exponential).
    \item Case (2) above where $i$ is linear (linear-linear).
    \item Case (2) above where $i$ is exponential (linear-exponential).
    \item Case (1) above where $i$ is linear (exponential-linear).
\end{enumerate}
If the increase in penalty $P_i$ from each case is at most $k$ times the optimal (either $E^i_{OPT}$ or $L^i_{OPT}$), then our algorithm is $k$-competitive. We analyze the above cases below.
\subsubsection*{Case 1: Exponential-Exponential}
\par We claim that the ordering of the exponential jobs is $\frac{2x-1}{x-1}$-competitive and formalize this with the \textbf{potential lemma}: we claim that at any point in time, the maximum potential of any exponential job in our ordering is at most the maximum potential of any job in the optimal ordering.
We will prove this by strong induction.
\begin{proof}
\par The base case is time $t=0$. Obviously the statement is true for this case.
\par For the inductive step, assume that at each time $t'\le t$ for some $t \ge 1$ that the statement is true. We want to prove it is true for $t+1$. Note at time $t$, our ordering always chooses the job with highest potential. Also note that when we choose to process a job $j$ at time $t$, $j$'s potential will not go up for that time step (since we decrement $t_j$ and multiply $s'_i$ by $x$, which offset each other).
\par Let the maximum potential job be $i$ in the optimal and $i'$ in our ordering with a potential of $T$.
\par There are two cases where the optimal's maximum potential job at time $t+1$ could be less than our ordering's maximum potential job.
\subsubsection*{Subcase 1}
\par Our ordering has at least 2 jobs $i',j'$ of potential $T$ and the optimal just has job $i$. The optimal schedules job $i$ during time $t$, so its maximum potential stays the same, but since our ordering can only schedule $1$ of $i'$ or $j'$, our ordering's potential goes up by a factor of $x$.
\par Assume this happens. Find $i'$ and $j'$ in the optimal ordering; one of them cannot be $i$. Without loss of generality, assume $j' \neq i$. Since we assumed $j'$ in the optimal ordering has potential less than $i$, this means that the optimal ordering processed $j'$ for at least 1 time unit more than our ordering. So at some $t'<t$, the optimal ordering processed $j'$ while we processed some other job $j'_2$. Why did we process $j'_2$? This is because, at the time, the potential of $j'_2$ was at least the potential of $j'$. This means had we not processed $j'_2$, it's potential would be at least $j'$'s potential at time $t$. Since the optimal ordering did not process $j'_2$ at the time, shouldn't $j'_2$'s potential time $t$ be at least $j$'s potential in our scheduling? The only way for this not to have happened is again for the optimal ordering to have processed $j'_2$ at least 1 more time unit than our ordering had as; say the optimal ordering processed $j'_2$ at time $t'_2$ and we processed some other job $j'_3$ at that time. By the same logic, we can infer that there is some time $t'_3$ such that the optimal ordering scheduled job $j'_3$ at that time and our ordering scheduled $j'_4$ instead; etc. If we repeat this $t$ times, and then we will have run out of time slots, so this situation is impossible to happen.
\subsubsection*{Subcase 2}
\par Our ordering has a job $i'$ of potential $T$ with greater than 1 unit of processing time remaining and the optimal has a job $i$ of potential $T$ with 1 unit of processing time remaining. The optimal schedules job $i$ during time $t$, completing it, so it's maximum potential drops, but since in our ordering  $i'$ is not done processing, our maximum potential stays the same.
\par First note $i' \neq i$ since they have the same potential but different processing times left. Thus, $i'$ in the optimal ordering has less $T$ potential, which means that it must have been processed for at least 1 more unit of time in the optimal ordering than in our ordering. Using the same logic as above, this leads to an infinite loop which eventually runs out of time slots, so this situation cannot happen.
\subsubsection*{Subcase 3}
\par Our ordering has a job $i'$ of potential $T$ with 1 unit of processing time remaining and the optimal has a job $i$ of potential $T$ with 1 unit of processing time remaining. We schedule $i'$ and the optimal schedules $i$, completing our respective jobs, but then our maximum potentials drop to different numbers. Say that our maximum potential is now held by job $j'$ and the optimal's maximum potential is held by job $j$, which is less than $j'$'s potential. However, since $j'$ in the optimal ordering has less than $j'$'s potential in our ordering, using the same logic as above, the optimal ordering had to have processed $j'$ 1 more time unit than our ordering, so we enter the infinite loop, meaning the situation is impossible. 
\par So either both our ordering and the optimal ordering have at least 2 jobs with $T$ potential, meaning at time step $t+1$, the maximum potential is $Tx$, or both our orderings have single incomplete jobs with $T$ potential, meaning the maximum potential stays the same, or both our jobs have single almost complete jobs with $T$ potential, and after completing them, the maximum potential drops to the same number. Thus, the maximum potential for the optimal order at time $t$ and $t+1$ is at least the maximum potential for our ordering.
\par By mathematical induction, we have proved the statement for all $t$.
\end{proof}

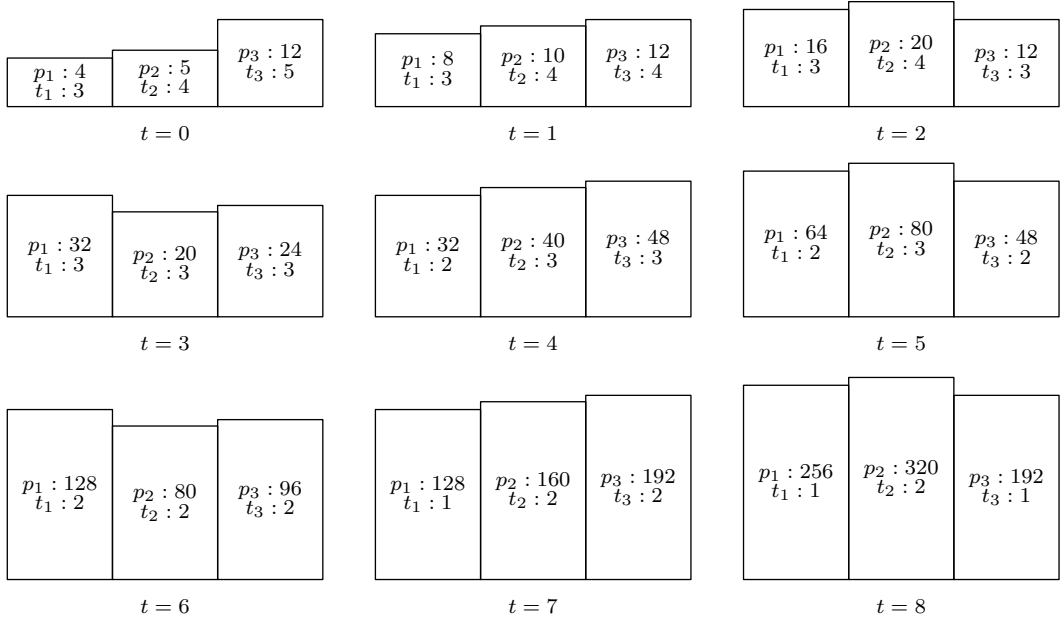
\begin{figure}[H]
    \centering
    \begin{asy}
        size(14cm);
        import graph;
        defaultpen(fontsize(8pt));
        
        void rct(real ap, real bp, real cp, real t1, real t2, real t3, real t, real yp){
            real f = 1.5;
            real a = log(ap)/f;
            real b = log(bp)/f;
            real c = log(cp)/f;
            real x = 7*(t
            real y = -yp;
            draw((x,y)--(x+2,y)--(x+2,a+y)--(x,a+y)--cycle);
            label(format("$p_1: 
            label(format("$t_1: 
            draw((x+2,y)--(x+4,y)--(x+4,b+y)--(x+2,b+y)--cycle);
            label(format("$p_2: 
            label(format("$t_2: 
            draw((x+4,y)--(x+6,y)--(x+6,c+y)--(x+4,c+y)--cycle);
            label(format("$p_3: 
            label(format("$t_3: 
            label(format("$t=
        }
        
        rct(4,5,12, 3,4,5, 0,0);
        rct(8,10,12, 3,4,4, 1,0);
        rct(16,20,12, 3,4,3, 2,0);
        rct(32,20,24, 3,3,3, 3,4);
        rct(32,40,48, 2,3,3, 4,4);
        rct(64,80,48, 2,3,2, 5,4);
        rct(128,80,96, 2,2,2, 6,9);
        rct(128,160,192, 1,2,2, 7,9);
        rct(256,320,192, 1,2,1, 8,9);
        //rct(512,320,384, 1,1,1, 9,15);
        //rct(512,640,768, 1,1,0, 10,15);
        //rct(1024,1280,768, 1,0,0, 11,15);
        //rct(2048,1280,768, 0,0,0, 12,21);
    \end{asy}
\end{figure}\begin{figure}[H]
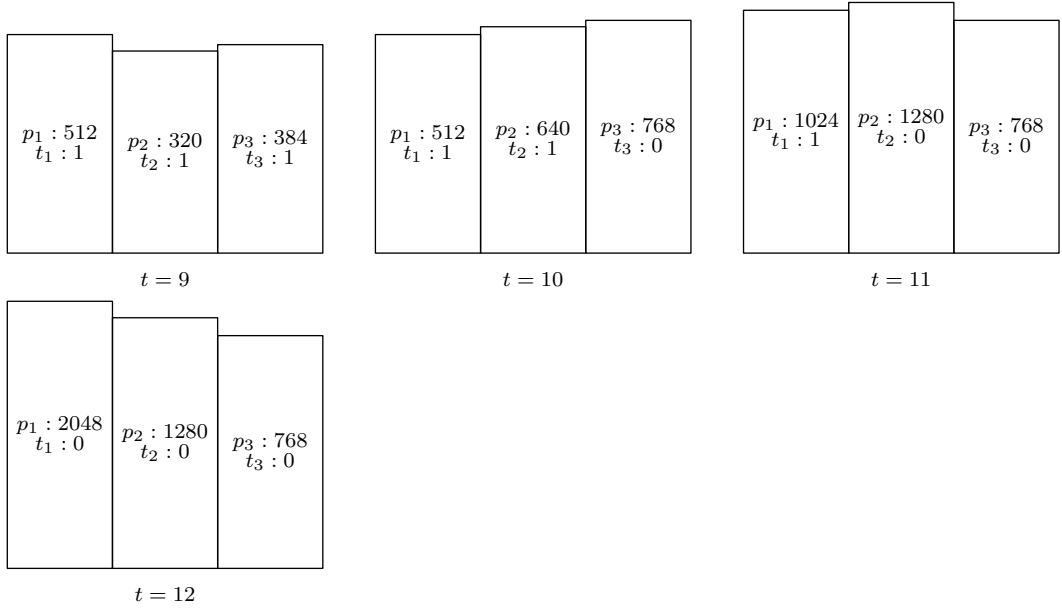

    \centering
    \begin{asy}
        size(14cm);
        import graph;
        defaultpen(fontsize(8pt));
        
        void rct(real ap, real bp, real cp, real t1, real t2, real t3, real t, real yp){
            real f = 1.5;
            real a = log(ap)/f;
            real b = log(bp)/f;
            real c = log(cp)/f;
            real x = 7*(t
            real y = -yp;
            draw((x,y)--(x+2,y)--(x+2,a+y)--(x,a+y)--cycle);
            label(format("$p_1: 
            label(format("$t_1: 
            draw((x+2,y)--(x+4,y)--(x+4,b+y)--(x+2,b+y)--cycle);
            label(format("$p_2: 
            label(format("$t_2: 
            draw((x+4,y)--(x+6,y)--(x+6,c+y)--(x+4,c+y)--cycle);
            label(format("$p_3: 
            label(format("$t_3: 
            label(format("$t=
        }
        
        //rct(4,5,12, 3,4,5, 0,0);
        //rct(8,10,12, 3,4,4, 1,0);
        //rct(16,20,12, 3,4,3, 2,0);
        //rct(32,20,24, 3,3,3, 3,4);
        //rct(32,40,48, 2,3,3, 4,4);
        //rct(64,80,48, 2,3,2, 5,4);
        //rct(128,80,96, 2,2,2, 6,9);
        //rct(128,160,192, 1,2,2, 7,9);
        //rct(256,320,192, 1,2,1, 8,9);
        rct(512,320,384, 1,1,1, 9,15);
        rct(512,640,768, 1,1,0, 10,15);
        rct(1024,1280,768, 1,0,0, 11,15);
        rct(2048,1280,768, 0,0,0, 12,21);
    \end{asy}
    \caption*{Example of cascading effect with three jobs for $x=2,n=3$}
\end{figure}
Let's consider the following example: we have $n$ identical exponential jobs (an example is provided above). In the optimal schedule, we will be scheduling them one by one until each finishes. However, in our policy, we will be continually preempting and doing a rotation (we process the job with the highest potential, so if we get to the point where all jobs have the same potential, we will sequentially cycle through the jobs and process each job for one second before switching to the next job in our cycle) of the $n$ jobs at every time step; eventually, if the processing times are long enough, all $n$ jobs will be within a $x$ factor of each other at each time step. This is because each job we are currently processing does not increase in potential while the other jobs in the rotation all increase by $x$ in potential. The current job we are processing will therefore rotate between the $n$ jobs. If we observe the ending behavior, in our optimal we  will be processing the last job while in our policy schedule we will have all $n$ jobs left with processing time 1. In our policy schedule, we will finish each of these n jobs one by one creating a cascading effect of jobs finishing. Note, the ending job in both the optimal schedule and our policy schedule will be the same job in terms of the total penalty incurred. The only difference is the $n-1$ jobs in our policy schedule which are finished in consecutive order immediately before the last job. The penalties incurred on these $n-1$ jobs will be $\sfrac{W}{x}, \sfrac{W}{x^2}, \sfrac{W}{x^3},\cdots$. Notice this creates a geometric series and therefore is upper bounded by $W\frac{x}{x-1}$. Because we always finish at the same time as the optimal, one job in the optimal has penalty $W$. Thus, in this example, our policy schedule is at least $\frac{W\frac{x}{x-1}}{W}=\frac{x}{x-1}$-competitive.
\par This cascading effect (where at the end the jobs all finish in consecutive units of time) happens for any group of jobs we are given, provided their processing times are long enough. For a group of jobs $j_1,j_2,\dots,j_k$, the algorithm will keep processing the one with the highest potential until all of the potentials of jobs are within an $x$ factor of each other. Then, it will rotate through the jobs until they finish an $x$ factor apart.
\par Let the jobs be sorted by increasing starting value, so $j_1$ has the lowest $s_1$ and $j_k$ has the highest $s_k$. Now, we will show this cascading effect is prevalent in any schedule and will always lead to a $(x+\frac{x}{x-1})$-competitive algorithm. Consider the graph where at each point the maximum potential is plotted. Notice that what happens is the job with the maximum potential rises by an $x$ factor every second until it starts processing, and when it is done processing, the maximum potential drops. We define the time where the job completes as a "peak." Note that the maximum potential may not actually drop, as there might be a second maximum that continues rising, so a peak can be contained within another peak. As we proved previously, the maximum potential job in the optimal schedule at every time step is greater than the maximum potential job in our ordering. We claim that each peak in our ordering corresponds to a unique peak in the optimal ordering.
\par Clearly, each peak in our ordering corresponds to a peak in the optimal, since the optimal always has at least as large of a maximum potential as our ordering. We will prove no two peaks in our ordering map to the same peak. Each group of cascading jobs in our policy will correspond to a group of jobs that are going to be done consecutively. We claim it is impossible to have multiple peaks in our policy that correspond to one peak in our optimal.
Let a group of jobs $j_1,j_2,\dots,j_k$ be defined as when our ordering processes $j_1,j_2,\dots,j_k$ in the rotating pattern showed above, where all potentials of these jobs are within an $x$ factor of each other. If all jobs in the group have equal remaining processing time, the group will end in a cascade: since we rotate through the jobs, they will finish on the same rotation, so if the last job that finishes has penalty $X$, the second to last will have penalty $\frac{X}{x}$, then $\frac{X}{x^2}$, etc. If some jobs in the group have less processing time, they will just drop out of the rotation earlier, so the total penalty will be less than $X+\frac{X}{x}+\frac{X}{x^2}+\frac{X}{x^3}+\dots$. Additionally, since when the rotation happens, the penalties between jobs in the same group can be at most an $x$ factor away from each other, the penalty is actually upper bounded by $X+x(\frac{X}{x}+\frac{X}{x^2}+\frac{X}{x^3}+\dots)$ (we don't multiply the first term by $x$ since it can not be greater than $X$ by the Potential Lemma). Notice first that our ordering processes jobs in groups, with each group corresponding to a peak (so the group of jobs that are being processed at a given moment are the largest potential active jobs); groups can be of a singular job, which happens when a job has way higher potential than any other job.
\par We will prove by induction that every group maps to a peak, and two groups mapping to the same peak with penalty $X$ can be adjusted to show that the sum of the penalties of the two groups will be at most $\O(X+X+\frac{X}{x}+\frac{X}{x^2}+\frac{X}{x^3}+\cdots) \equiv \O(X+X\frac{x}{x-1})\equiv \O(\frac{2x-1}{x-1}X)$.
\begin{proof}
    We first prove the base case: we show the first group follows this property. 
    If the optimal finishes processing all the jobs in group $1$ before all the jobs in group 2, this means that group $1$ corresponds to a unique peak in both the optimal and our ordering, because group 2 in our ordering will then correspond to the unfinished jobs in group 2 of the optimal ordering.
    \par We claim this immediately shows the property holds. We know the peak of group 1 in the optimal ordering is at least the peak of group 1 in our ordering by the Potential Lemma. Let the peak of the optimal ordering be $X$ penalty, so our ordering's peak is at most $X$. As said before, group 1 will have total penalty upper bounded by $X+X+\frac{X}{x}+\frac{X}{x}+\dots$.
    \par Now consider the case where the optimal ordering finishes group $1$'s jobs after group 2's jobs. Say that the last job scheduled in group $1$ is $o'$. We will analyze how this affects the peaks. Say we need to schedule jobs $j_1,j_2,\dots,j_k$, ordered by current penalty. Which job should we schedule last so that the ending penalty of the last job scheduled is minimized? No matter the order we schedule, the last job always ends at the same time, so we should schedule the job with the least current penalty last: $j_1$. We claim that each of the current penalties of $o_1,o_2,\dots,o_k$ are greater than each of $p_1,p_2,\dots,p_k$. Assume there is some $p'$ with current penalty $s'_{p'}$ greater than a job $o'$. Since in our ordering $p'$ completes after $o'$ completes, we know that for every unit of time before $o'$ completes, $p'$'s potential is at most $o'$'s potential. Since potential is $s'_ix^{t_i}$, $p'$'s current processing time then must be less than $o$'s processing time, $t_{p'}<t_{o'}$. But since we will eventually schedule $o'$ to completion, there will be a time in the future such that $t_{p'}=t_{o'}$, and since we still schedule $o'$ before $p'$, then it means at that moment $o'$'s potential is greater than $p'$, so $s'_{o'}x^{t_{p'}}>s'_{p'}x^{t_{p'}} \longleftrightarrow s_{o'}>s'_{p'}$, a contradiction.
    \par This same logic follows for any two jobs $i$ and $j$ such that $i$ is scheduled before $j$ in our scheduling (given that they are both released before one of them completes), so our ordering therefore schedules the minimum penalty job last, so it has the minimum last penalty. Since we assumed the optimal ordering schedules a job in group $1$ last, this means its ending penalty (peak) $X$ is larger than our peak in group $2$. Note that in our current schedule, by the time all of group 1 is finished processing, all jobs in group 2 have less potential than the last job in group $1$ (otherwise those jobs would just be in group 1).
    \par  Now imagine we swapped: we scheduled group 2 before group 1. Since group 1 follows the cascading pattern, our penalty for group 1 would be at most $X+X+\frac{X}{x}+\dots$. Since we showed that group 2's penalty was so low, the penalty of the last job finished in group 2 now is less than the first job finished in group 1. Thus, all of group 2's penalties can be included in the $X+X+\frac{X}{x}+\dots$ series, so in this case the sum of both groups is at most $\frac{2x-1}{x-1}$ times the optimal's peak $X$.
\end{proof}
\par We will claim that the total penalty incurred by both group 1 and group 2 is not greater than the penalty incurred by processing group 2 first and then group 1.
\begin{proof}
\par We consider two scheduled groups and claim that swapping the groups (switching the ordering of them) does not decrease the sum of penalties. We recall that sequentially-completed elements within each group differ by at most $x-1$. We denote the number of jobs in the first group $n_1$, the number of jobs of the second group $n_2$ and we number sequentially-completed jobs from $1$ to $n_2+n_1$. We analyze using the initial penalty $s_{n_1}$ of the last job in the first group, which we denote $y$ for ease of typing. We also note that the penalty at the completion of the first job in the second group is less than the penalty at the completion of the last job in the first group; formally, $s'_{n_1}>s'_{n_1+1}$. We show that swapping two groups does not decrease the sum of penalties by considering the worst case when swapping: in order to minimize the difference between the sum of the second group post- and pre-swap, we maximize the penalty sum of the second group. Likewise, in order to minimize the difference between the sum of the first group post- and pre-swap, we minimize the penalty sum of the first group.
\par We minimize the difference between the sum of the first group's penalties post- minus pre-swap by setting the singular element ($n_1=1$) in the first group to $s_1=y$ (we have a single element in the first group because this minimizes the sum of penalties in the first group, minimizing the difference between the sum of the second group across group swapping). Likewise, to minimize the difference between the sum of the second group across group swapping, we maximize the sum of the second group. We can do this for arbitrary $n_2$ by setting $s_2=\frac{y_1-\eps_1}{x},\eps_1>0$, then increasing each sequentially-completed job's potential $s'_i$ by $x$ (this maximizes the sum of the second group while ensuring every job in the second group is not greater than a factor of $x$ from its adjacent jobs. This results in an initial sum of group penalties of
\begin{equation*}
    \underbrace{yx^1}_{\mathclap{\text{first group}}} + \underbrace{\sum_{i=2}^{n_2+1}{\frac{y-\eps}{x}x^{i}}}_{\mathclap{\text{second group}}}
\end{equation*}
The sum of group penalties after swapping the groups is then
\begin{equation*}
    \underbrace{x^{k+n_2}yx^1}_{\mathclap{\text{first group}}} + \underbrace{x^{-k-n_2}\sum_{i=2}^{n_2+1}{\frac{y-\eps}{x}x^{i}}}_{\mathclap{\text{second group}}}
\end{equation*}
where $k\geq 0$ ($n_2+k$ is the number of timesteps between after the last job in the first peak is processed and the last job in the second peak is processed). We see that to minimize the post-swap sum, we set $k=0$. This yields
\begin{equation*}
    \underbrace{x^{n_2}yx^1}_{\mathclap{\text{first group}}} + \underbrace{x^{-n_2}\sum_{i=2}^{n_2+1}{\frac{y-\eps}{x}x^{i}}}_{\mathclap{\text{second group}}}
\end{equation*}
However, we know that
\begin{equation*}
    yx^1+\sum_{i=2}^{n_2+1}{\frac{y-\eps}{x}x^{i}} = \frac{y-\eps}{x}x^{n_2+2}+\eps x < yx^{n_2+1}\text{ and }x^{n_2}yx^1 = \underbrace{yx^{n_2+1}}_{\mathclap{\substack{\text{first group}\\ \text{postswap}}}}
\end{equation*}
\begin{equation*}
    \Longrightarrow yx^{n_2+1} > \underbrace{yx^1}_{\mathclap{\text{first group}}} + \underbrace{\sum_{i=2}^{n_2+1}{\frac{y-\eps}{x}x^{i}}}_{\mathclap{\text{second group}}}
\end{equation*}
Therefore, the sum of penalties in the first group alone post-swap is greater than the sum of penalties in both groups pre-swap and because we originally chose to maximize the sum of penalties in the second group and minimize the sum of penalties in the first group between the swap, 
the sum of penalties post-swap even when trying to minimize this sum is greater than the sum of group penalties pre-swap. As a result, swapping groups cannot decrease the penalty sum; in fact, swapping groups will increase the penalty sum. 
\end{proof}
Thus, for the two-group case, our algorithm's ordering's total penalty is at most $\frac{2x-1}{x-1}$ times the optimal's peak.

We can easily extend this to 3 or more peaks: whenever the optimal ordering finishes a job in group 1 after finishes all jobs in groups $1,2,\dots,k$ for some $k>1$, the bound still holds.
\par Assume that for some $n$, the first $n$ groups follow this property. Showing it is true for $n+1$ is exactly the same as the base case. Thus, for each of the optimal ordering's peaks, we can group a set of peaks in our ordering such that the total cost of the group is at most $\frac{2x-1}{x-1}$ times the optimal ordering's peak penalty cost. Since each group in our ordering will be assigned to a peak in the optimal ordering, our ordering is then $\frac{2x-1}{x-1}$-competitive.
\subsubsection*{Case 2: Linear-Linear}
\par We claim that this ordering is $n$-competitive. At any point in time $t$, we claim the maximum weight of a still unfinished job in this ordering is less than or equal to the maximum weight of any unfinished job in the optimal ordering.\\ This is clearly true because we always process the greatest weight job. Thus, at any point in time, let the still unfinished linear jobs be $i_1,\dots,i_k$. Without loss of generality, say we are processing job $i_k$. Each of the other jobs has weight less than $w_{i_k}$. As claimed before, at time $t$, the optimal scheduling must have a job with at least $w_{i_k}$ weight. Thus, in the optimal, the penalty increase resulting from the unfinished jobs at time $t$ is at least $w_{i_k}$. The penalty increase at time $t$ for our scheduling is $\sum_{j=1}^kw_{i_j}$. If we divide our penalty incurred by the penalty incurred in the optimal scheduling, we get,
\begin{align*}
    \frac{\sum_{j=1}^kw_{i_j}}{w_{i_k}}&\le \frac{\sum_{j=1}^kw_{i_k}}{w_{i_j}}\\
    &=k\\
    &\le n
\end{align*} 
Thus, this ordering is $n$ competitive.
\subsubsection*{Case 3: Exponential-Linear}
\par Say an exponential job $i$ is processed before some set of linear jobs $l_1,l_2,\cdots,l_k$ but in the optimal scheduling these linear jobs are processed before $i$.
\par The extra penalty we incur is equal to the time $t_i$ the linear jobs have to wait for $i$ to finish multiplied by the sum of the weights of all the linear jobs. To get the competitiveness of this, we divide by the optimal penalties of these jobs, which is scheduling $i$ after the linear jobs.
We get
\begin{align*}
    \frac{t_i(w_{l_1}+\dots+w_{l_k})}{s'_ix^{t_i+t_{l_1}+\dots+t_{l_k}}+\sum_{j=1}^k{w_j}}&\le \frac{t_i}{x^{t_i}}\frac{M}{s_i}\frac{k}{x^k}
\end{align*}
Clearly, this ratio is maximized when $w_{l_j}=M$ and $t_{l_j}=1$ for each $1\le j \le k$. The fraction becomes
\begin{align*}
    \frac{t_ikM}{s'_ix^{t_i+k}+kM}
\end{align*}
If $t_i=\log{\frac{Mk}{s'_i}}+\alpha$ for some $\alpha>0$ (note this $\alpha$ does not refer to the one used in $\alpha$-point analysis and is a constant meant to make $t_i>\log{\frac{Mk}{s'_i}}$), $s'_ix^{t_i+k}>x^kMk$, then the fraction becomes,
\begin{align*}
    \frac{t_ikM}{s'_ix^{t_i+k}+kM}
    &\le \frac{(\log(\frac{Mk}{s'_i})+\alpha)kM}{x^{k+\alpha}Mk}\\
    &\le \frac{(\log(\frac{Mk}{s'_i})+\alpha)}{x^{k+\alpha}}\\
    &\le \frac{\log(\frac{Mk}{s'_i})}{x^{k+\alpha}}+\frac{\alpha}{x^{k+\alpha}}
\end{align*}
$\frac{\alpha}{x^{k+\alpha}}$ goes to 0 as $\alpha$ increases (this term can never be greater than 1 because $x\geq 2$), so in this case the operation is $\log{\frac{Mk}{s'_i}}$ competitive (it is actually $(\log{\frac{Mk}{s'_i}}+1)$-competitive, but the $+1$ is so small relative to $\log{\frac{Mk}{s'_i}}$ that we ignore it in our analysis). 
\\If $t_i<\log{\frac{Mk}{s'_i}}$, we have
 \begin{align*}
    \frac{t_i`kM}{s'_ix^{t_i+k}+kM}&\le \frac{t_ikM}{kM}\\
    &\le t_i\\
    &\le \log{\frac{Mk}{s'_i}}
\end{align*}
Thus, in both cases, this operation is $\log{\frac{Mk}{s'_i}}$ competitive with the optimal. However, in the worst case, will need to do this for every exponential job, and the competitiveness is additive since the $l_1,l_2,\dots,l_k$ of 1 exponential job may overlap with another exponential job's linear set. Thus, the operation is overall $n\log{\frac{Mk}{s'_i}}$ competitive.
\subsubsection*{Case 4: Linear-Exponential}
\par An exponential job $i$ is processed after some set of linear jobs $l_1,l_2,\dots,l_k$ but in the optimal scheduling these linear jobs are processed after $i$. We will compare $i$'s penalty with the penalty of $i$ in our ordering.
\par We claim that at the time job $i$ is released, when the current penalty of $i$, $s'_i$, becomes more than $\sqrt{M}$ (recall $s'_i$ starts at $s_i$ and multiplies by $x$ each time unit), no linear jobs will be scheduled before $i$.
\par Assume for contradiction that this is not true, so after being on the queue at least $\log{\sqrt{\frac{M}{s_{j}}}}$ units of time, a linear job is still being scheduled before $i$. At that point in time, we know if $i$ is at the top of the queue, it will be processed instead of any linear job. So there must be a job $j$ at the top of the queue with $t_j<\log{\sqrt{\frac{M}{s'_{j}}}}$. In order for $j$ to be before $i$ in the priority queue, we must have that $\frac{s'_ix^{t_i}}{x^{t_i}-1}<\frac{s'_jx^{t_j}}{x^{t_j}-1}$. Substituting $\sqrt{M}$ for $s'_j$, we have,
\begin{align*}
    \frac{s'_ix^{t_i}}{x^{t_i}-1}&<\frac{s'_jx^{t_j}}{x^{t_j}-1}\\
    \Longleftrightarrow \ \ \ \ \ \ \ \ \ \frac{s'_i\sqrt{\frac{M}{s'_i}t}x^{t_i}}{x^{t_i}-1} &<\frac{s'_jx^{t_j}}{x^{t_j}-1}\\
     \Longleftrightarrow \ \ \ \ \ \ \ \ \ \ \frac{\sqrt{s_{\min}}\sqrt{M}x^{t_i}}{x^{t_i}-1} &<\frac{s'_jx^{t_j}}{x^{t_j}-1}\\
    \Longleftrightarrow \ \ \ \ \ \ \ \ \  \sqrt{s_{\min}}\sqrt{M}&<s'_j\frac{x^{t_j}}{x^{t_j}-1}\\
    \Longleftrightarrow \ \ \ \ \ \ \ \ \  \sqrt{s_{\min}}\sqrt{M}&<s'_jx^{t_j} \ \ \ \ \ \ \ \ \ \ \ \text{(since $x^{t_j}-1 \ge 1$)}\\
    \Longleftrightarrow \ \ \ \ \ \ \ \ \  \sqrt{\frac{M}{s'_j}}&<x^{t_j} \ \ \ \ \ \ \ \ \ \ \ \text{(since $x^{t_j}-1 \ge 1$)}\\
    \Longleftrightarrow \ \ \ \ \ \ \ \ \ \ \log\sqrt{\frac{M}{s'_j}}&<t_j \ \
\end{align*}
Which, by the definition of our algorithm, means that $j$ will be scheduled before any linear job, which is a contradiction.
\par Thus, $i$ will stop waiting for any linear jobs to schedule after $s'_i$ has reached at least $\sqrt{\frac{M}{s}}$. So $i$'s penalty will be at most $x^{\log\sqrt{\frac{M}{s_{i}}}}=\sqrt{\frac{M}{s_{i}}}$ times larger than its penalty in our exponential ordering. Thus, $P^j$ is $\sqrt{\frac{M}{s_{i}}}$-competitive with our exponential ordering.
\par For each exponential job $i$, by case 1 and case 4, its cost is $\dfrac{2x-1}{x-1}\sqrt{\frac{M}{s_{i}}}E^i_{OPT}$, and by case 3, we will charge an extra $n\log(\frac{Mn}{s_{\min}})E^i_{OPT}$ for any linear penalties misscheduled after it. If this operation is $X$ competitive, we can solve for $X$ as follows,
\begin{align*}
X&=\frac{\dfrac{2x-1}{x-1}\sqrt{\frac{M}{s_{i}}}E^i_{OPT}+n\log(\frac{Mn}{s_{\min}})E^i_{OPT}}{E^i_{OPT}}\\
&=\sqrt{\frac{M}{s_{\min}}}\dfrac{2x-1}{x-1}+n\log\left(\frac{Mn}{s_{\min}}\right)\\
\end{align*}
For each linear job $i$, by case 2, its cost is $nL^i_{OPT}$ compared to all linear jobs, and note that in case 3, we charged the extra cost incurred by waiting for exponential jobs to the exponential jobs, so this adds no extra cost for $i$. Thus, this operation is $\frac{nL^i_{OPT}}{L^{i}_{OPT}} =n$ competitive.\\
The total penalty of our scheduling is $\sum_{i \in E} P_i + \sum_{i \in L} P_i$. We know,
\begin{align*}
    \sum_{i \in E} P_i + \sum_{i \in L} P_i&=\sum_{i \in E} \left[ \dfrac{2x-1}{x-1}\sqrt{\frac{M}{s_{\min}}}E^{i}_{OPT} + \log\left(\frac{Mn}{s_{\min}}\right)\right]+\sum_{i \in L} nL^{i}_{OPT}\\
    &\le \left(\dfrac{2x-1}{x-1}\sqrt{\frac{M}{s_{\min}}} + n\log\left(\frac{Mn}{s_{\min}}\right)\right)\left(\sum_{i \in E} E^{i}_{OPT} +\sum_{i \in L} L^{i}_{OPT}\right)\\
    &=n\left(\sqrt{\frac{M}{s_{\min}}} + \log\left(\frac{Mn}{s_{\min}}\right)\right)C_{OPT}
\end{align*}
So the new scheduling is $\left(\dfrac{2x-1}{x-1}\sqrt{\frac{M}{s_{\min}}} + n\log\left(\frac{Mn}{s_{\min}}\right)\right)$ competitive. Note that $\dfrac{2x-1}{x-1} \leq 4$ if we assume $x \geq 2$ and approaches 2 as $x$ goes to infinity. Assuming the number of jobs is much less than the maximum penalty and the  base $x$ is sufficiently large, we have improved from the naive algorithm by a $\dfrac{\sqrt{\frac{M}{s_{\min}}}}{4}$ factor. However, note that the bottleneck is in case 4, which is when we schedule linear jobs before exponential jobs.
\section{A Minor Modification to Our Algorithm}
Let us consider the policy of always scheduling linear jobs after exponential jobs, but keeping the same sorting order of linear jobs and exponential jobs.
\par Cases 1 and 2 are unchanged, and case 4 now doesn't add any extra penalty. For case 3, the analysis still applies. So we now have a $n\log(\frac{Mn}{s_{\min}})$-competitive algorithm. Additionally, note that we no longer need to make the problem partially online anymore since the algorithm does not use $M$ or $s_{\min}$.
\section{Conclusion \& Further Research}
\par We discuss the scheduling problem of jobs with either exponential or linear penalties with preemption by reviewing papers on related scheduling problems and presenting an online competitive algorithm for the problem. We review two papers and note that existing methods to create and analyze a competitive algorithm for scheduling problems include speed augmentation, which compares an algorithm on machines running at full speed to the optimal schedule on machines running $1+\eps$ slower than full speed, and $\alpha$-points, which measure the point at which $\alpha\in (0,1]$ of a job has completed and use these points to compare a schedule to the optimal. We present these reviews in our paper as examples of methods of analyzing competitiveness. Finally, we describe and analyze the competitiveness of a novel algorithm for job scheduling on exponential/linear jobs with preempting, showing that our algorithm is 
asymptotically $n\sqrt{\frac{M}{s_{\min}}} + \log(\frac{Mn}{s_{\min}})$
-competitive with the optimal and better than the na{\" i}ve algorithm, and then we made a final modification that results in an algorithm that is $n\log{\frac{Mn}{s_{\min}}}$-competitive with the optimal. \\
In the process of proving the competitiveness of these algorithms, we proved the constant time competitiveness of our job ordering using only exponential jobs, which was a very interesting result. Note that our bounds could be improved with more precise competitive analysis: our analysis merely had the goal of beating the na{\" i}ve algorithm by a nonconstant margin. Further research might be able to be done to remove the log factor as well. This could have implications for improving OS scheduling algorithms and could be expanded to include multiple machines to simulate multicore job processing.
\newpage
\bibliographystyle{unsrt}
\bibliography{refs} 
\end{document}